\documentclass[aps,prl,twocolumn,showpacs,psfig,superscriptaddress,longbibliography]{revtex4-1}
\usepackage{textcomp}
\usepackage{times}
\usepackage{graphicx}
\usepackage{float}
\usepackage{latexsym,amsmath,amssymb,bm,euscript}
\usepackage{color}
\usepackage{subfigure}
\usepackage{epstopdf}
\usepackage[colorlinks=true,linkcolor=blue,citecolor=blue,urlcolor=blue]{hyperref}
\usepackage{hyperref}
\usepackage{soul}
\usepackage[normalem]{ulem}
\usepackage{mathrsfs}
\usepackage{amsmath,amssymb}
\usepackage{lettrine}
\usepackage{xfrac}
\usepackage{xspace}
\usepackage{textcomp}
\usepackage{wasysym}
\usepackage{xr}
\graphicspath{{./Figs/}}

\begin{document}

\title{Interaction-driven roton condensation in a $C=2/3$ fractional quantum anomalous Hall state}
\author{Hongyu Lu}
\affiliation{Department of Physics and HKU-UCAS Joint Institute
	of Theoretical and Computational Physics, The University of Hong Kong,
	Pokfulam Road, Hong Kong SAR, China}
\author{Han-Qing Wu}
\affiliation{Guangdong Provincial Key Laboratory of Magnetoelectric Physics and Devices, School of Physics, Sun Yat-sen University, Guangzhou 510275, China }
\author{Bin-Bin Chen}
\affiliation{Department of Physics and HKU-UCAS Joint Institute
	of Theoretical and Computational Physics, The University of Hong Kong,
	Pokfulam Road, Hong Kong SAR, China}
\author{Kai Sun}
\email{sunkai@umich.edu}
\affiliation{Department of Physics, University of Michigan, Ann Arbor, Michigan 48109, USA}
\author{Zi Yang Meng}
\email{zymeng@hku.hk}
\affiliation{Department of Physics and HKU-UCAS Joint Institute
	of Theoretical and Computational Physics, The University of Hong Kong,
	Pokfulam Road, Hong Kong SAR, China}

\begin{abstract}
The interplay of topological order and charge order exhibits rich physics. Recent experiments that succesfully realized the frational quantum anomalous Hall (FQAH) effect in twisted  MoTe$_2$ bilayers and rhombohedral multilayer graphene without external magnetic field further call for deeper understanding of the relation between topological order and charge order in quantum moir\'e materials. In the archetypal correlated flat-band model on checkerboard lattice, a FQAH smectic state with coexistent topological order and smectic charge order has been numerically discovered at filling $\nu=2/3$~\cite{LHY2024FQAHS}.
In this work, we explore the global ground-state phase diagram of the model with competing interactions and find a $C=2/3$ FQAH phase surrounded by four different charge density wave (CDW) phases.
In particular, we identify a FQAH-CDW transition triggered by 
roton condensation, in that, the minimal roton gap continues to decrease at the same finite momentum, along with the diverging density flucuations at the transition point, after which the system enters into a CDW metal phase with the same ordered wavevector. Our discovery points out that the charge-neutral roton modes can play a significant role in a transition from FQAH topological order to CDW symmetry-breaking order, discussed in FQH literature while severely neglected in FQAH systems.

\end{abstract}

\date{\today }
\maketitle

\noindent{\textcolor{blue}{\it Introduction.}---} 
The states of matter, are by and large categorised according to either the topological or the Landau-Ginzburg symmetry breaking paradigms, and usually the states with topological order and Landau order do not interfere. But interesting phenomena happen when they do interfere, and presently the frontier of such interference lies at the phase transitions between fractional quantum (anomalous) Hall (FQH and FQAH) states and charge density wave (CDW) states~\cite{BYang2012bandmass,DNSheng2011boson, CDGong2020boson, Barkeshli2015_fci_continuous, WZhu2016kagomeboson,Valentin2023_fci_cdw, songPhase2023, XYSong2023fqahc, Reddy2023_fqah,nicholas2023_pressure_fci, shengQuantum2024}. 
Recent experiments that successfully identified the FQAH states in twisted MoTe$_2$ bilayers and pentaplayer graphenes~\cite{Heqiu2021_fci, LFu2021_fci, caiSignature2023,park2023_fqah,zengThermodynamic2023,xu2023_fci, kang2024fqsh,multilayer_graphene_fqah} manifest the importance of properly addressing the competition and transition between topological orders and CDW orders~\cite{Valentin2023_fci_cdw, songPhase2023, XYSong2023fqahc,xu2023maximally,yu2023fractional,Reddy2023_fqah, nicholas2023_pressure_fci,shengQuantum2024}.

In FQH literature, the magneto-roton modes are identified as the elementary low-energy collective excitations~\cite{Haldane1985_roton, FCZhang1985_impurity, Rezayi1985_impurity, GMP_SMA1985, girvinMagneto1986,kangInelastic2000,kukushkinDispersion2009,  Haldane2022_roton,KYang2022_roton}, and the possible transition from FQH to Wigner crystal via softening of the neutral collective modes has been widely discussed~\cite{Ho_BEC2001, Cooper_BEC2001, Sinha_rotonBEC2005, BYang2012bandmass,  Mukherjee2022_rotoncondense}. 
However, the study of roton condensates encounters a notable impediment: due to symmetry reasons, the softening of roton inevitably leads to a first-order transition, making it impossible to access the quantum critical point associated with such condensates. This obstacle arises from the isotropy of Landau levels, where low-energy roton excitations form an isotropic ring in momentum space. As these excitations soften, the energy of the entire 
ring approaches zero, contrasting with traditional Bose-Einstein condensates that feature energy minima at a single or few distinct momentum points. 
As documented in Brazovskii's seminal work~\cite{Brazovskii1975}, the Ginzburg-Landau free energy for such a system encompasses a cubic term, $\Psi_{\mathbf{k}_1}\Psi_{\mathbf{k}_2}\Psi_{\mathbf{k}_3}$, where $\Psi_{\mathbf{k}_i}$ is the order parameter with momentum $\mathbf{k}_i$ and $i=1,2, 3$ indicates three wavevectors 120 degrees apart. This cubic term implies that the transition will be first order~\footnote{we note similar mechanism dedicate the first order transition in frustrated transverse field Ising model on honeycomb lattice~\cite{wangCaution2017}}. Even in the absence of the cubic term, critical fluctuations will flip the sign of the vertex function to negative, making the transition first order~\cite{Brazovskii1975}.  
Consequently, the roton energy cannot be reduced to zero without interrupted by the instability induced by this cubic term, which  blocks the access to the quantum critical point of roton condensate.

In contrast to Landau levels, quantum critical points for roton condensates are permissible in FQAH systems. While FQAH states exhibit similarities to FQH states, including their ground state topology and excitations, they  begin to diverge regarding roton condensates: FQAH systems are governed by lattice space group symmetry rather than continuous space rotational/translational symmetry, which exempts them from the restrictions implied by Brazovskii's conclusions. This exemption allows for the possibility of a second-order phase transition of roton condensates, where roton energy reduces to zero and the CDW susceptibility diverges to infinity. Furthermore, FQAH systems can also support other novel phenomena unattainable in FQH systems. In Landau levels, CDW wavevectors are typically dictated by the filling factor, which generally results in incompressible CDW states. However, in FQAH systems, the CDW wavevector is influenced by both the filling factor and the lattice vector. As a result, the competition between these two length scales could stabilize compressible CDW states, which are energetically disfavored in Landau-level systems but are conceivable in FQAH systems, potentially facilitating quantum criticality between an FQAH state and a compressible CDW state via roton condensate.

Despite its theoretical plausibility, to date, there has been no experimental or model realization of this novel quantum critical point. The existing literature on FQAH systems delved into the analysis of the $C=1/2$ FCI's roton excitations using the single-mode approximation (SMA) after lowest-band projection~\cite{SMA_FCI2014}, studied the charge oscillation features~\cite{HYL2023_thermoFQAH,song2023halo} and the thermodynamic nature of roton modes~\cite{HYL2023_thermoFQAH, LHY2024FQAHS}. Theoretical proposals exist for transitions in FQAH systems, including those to CDW states~\cite{songPhase2023}; however, numerical evidence for roton-condensation-induced transitions remains elusive. This key disparity between theoretical arguments and numerical/experimental evidence raises a critical question: are these theoretical expectations overlooking some vital factor that precludes such quantum criticality, or is the principle sound, and the appropriate system to observe such phenomena has simply not yet been identified? Our findings suggest the latter is true.

In this work, we identify such a roton-driven transition by employing exact diagonalization (ED) and density matrix renormalization group (DMRG) simulations to study the archetypal correlated flat-band model on the checkerboard lattice at fractional filling $\nu=2/3$~\cite{Sun2011_fci, DNSheng2011_fci, Titus2011_fci, Bernevig2011_fci, XGWen2011_fci}. Previuosly, a FQAH smectic state of coexisting topological and smectic charge orders has been found~\cite{LHY2024FQAHS}, 
here we reveal that this model, in a broader phase space, provides the desirable platform for investigating the rich interplay betweem CDW and FQAH states. 
 Our main results include: i) the global ground state phase diagram of the model, in which a $C=2/3$ FQAH state without coexisting charge order is found to be surrounded by four CDW states with various charge orders; 
 ii) among the FQAH-CDW transitions, one of them is particular interesting, as during the transition 
 the lowest roton mode of the FQAH state at $(\pi,\pi)$ goes soft with closing gap 
 and finally condense into the ground state of a ($\pi,\pi$) CDW metal; 
 iii) the neutral charge fluctuation at $(\pi,\pi)$ at the transition point tends to diverge and such hidden information of excited states can be identified from the ground-state results of density-density correlations, largely overlooked in previous literatures.

\begin{figure}[htp!]
	\centering		
	\includegraphics[width=0.5\textwidth]{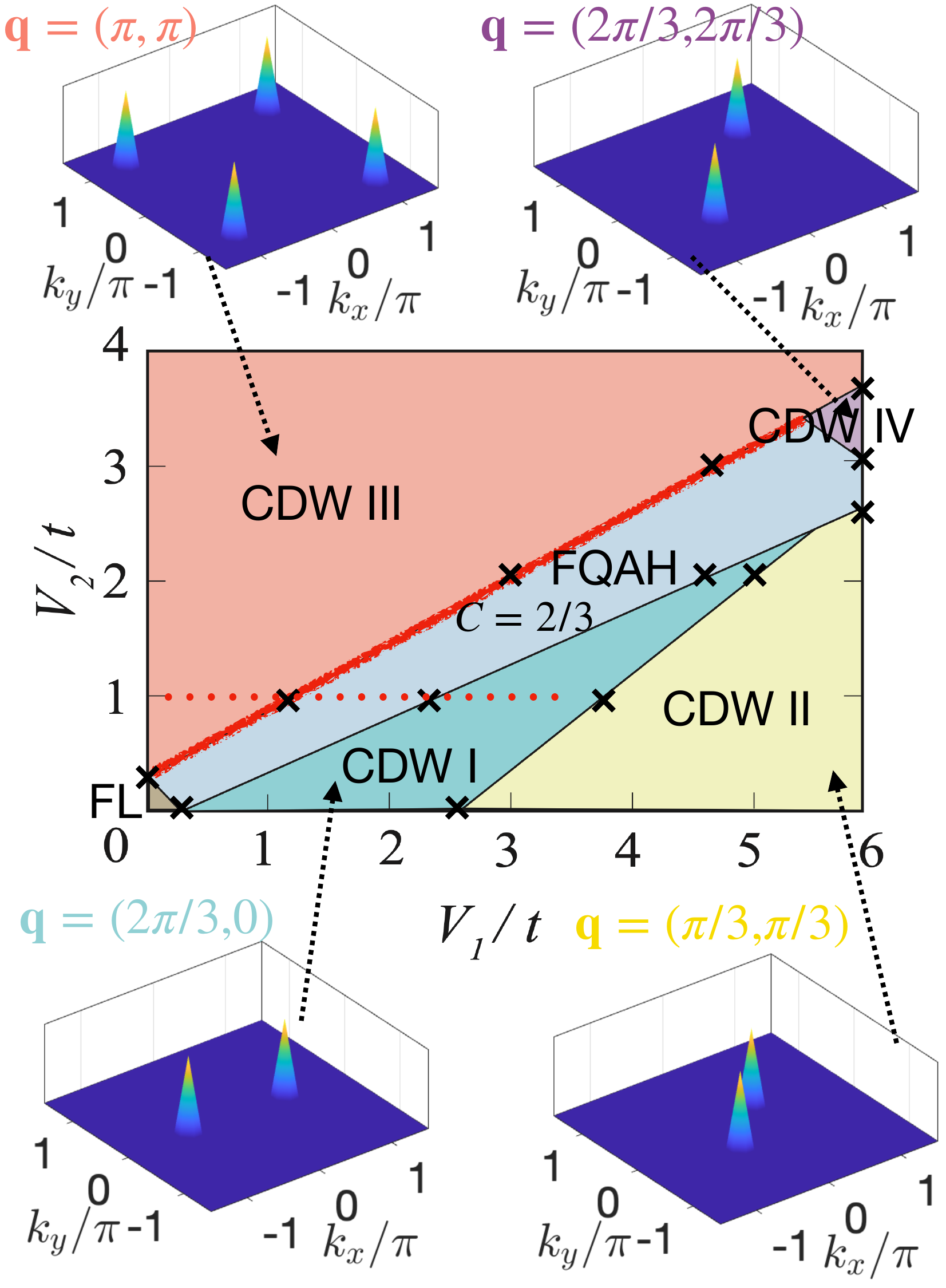}
	\caption{ \textbf{Global $V_1$-$V_2$ phase diagram.}  In the finite $V_1$-$V_2$ region, we find a robust $C=2/3$ FQAH state without CDW order while its roton minimum is at $(\pi,\pi)$. The FQAH state is surrounded by various topologically trivial charge-ordered phases named CDW I-IV with distinct ordered wave vectors specified. The schematic plots of $n(\mathbf{q})$ for CDW I-IV are shown respectively. When $V_1$ and $V_2$ are small, the ground state is a FL. The black crosses refer to the transition point determined from DMRG simulations and the black arrows connect the $n(\mathbf{q})$ plots to the phase diagram. We focus on the roton triggered transition between FQAH and CDW III, whose phase boundary is marked red, while we focus on the red dotted path in the main text.}
	\label{fig_phasediagram}
\end{figure}

\noindent{\textcolor{blue}{\it Model and methods.}---} We consider the $\nu=2/3$ filling of the flat band with spinless fermions on the checkerboard lattice~\cite{LHY2024FQAHS},
\begin{equation}
	\begin{aligned}
		H =&-\sum_{\langle i,j\rangle}te^{i\phi_{ij}}(c_i^\dagger c^{\ }_j+h.c.)-\sum_{\langle\hskip-.5mm\langle i,j \rangle\hskip-.5mm\rangle}t'_{ij}(c_i^\dagger c^{\ }_j+h.c.)\\
		&-\sum_{\langle\hskip-.5mm\langle\hskip-.5mm\langle i,j \rangle\hskip-.5mm\rangle\hskip-.5mm\rangle} t''(c_i^\dagger c^{\ }_j+h.c.)+\sum_{i,j}V(n_i-\frac{1}{2})(n_j-\frac{1}{2}),
	\end{aligned}
	\label{eq:eq1}
\end{equation}
with dimensionless parameters: : $t=1$ (as the energy unit), $t'_{ij}=\pm 1/(2+\sqrt{2})$ with alternating sign in edge-sharing plaquettes, $t''=-1/(2+2\sqrt{2})$ and $\phi_{ij}=\frac{\pi}{4}$. We consider the nearest-neighbor (NN) interaction $V_1$ and the next-nearest-neighbor (NNN) interaction $V_2$. We note in our previous work, only the third neighbor interaction $V_3$ has been included~\cite{LHY2024FQAHS}.

Our DMRG simulations utilize cylinders with periodic $N_y=4,\ 6$ and open $N_x$ up to 24, for the two-site unit cells. The total number of lattice sites $N=N_y\times N_x \times 2$, the average density $\bar n$ and the filling of the flat band $\nu$ are defined as $\bar n=N_e/N = \nu/2$. The DMRG simulations are based on the QSpace library~\cite{AW2012_QSpace} with charge U(1) symmetry and complex wavefunctions. We keep the maximum bond dimension up to $D=2048$ with maximum truncation error $\delta<10^{-5}$. The ED simulations are conducted up to a 36-site torus. We have used two geometries in ED: one is similar to DMRG where the unit cell consists of two sites, while in the other case, every four NN sites are repeated periodically, since we need different momentum sectors for the CDW states at finite sizes. More numerical details are shown in Supplemental Materials (SM)~\cite{suppl}.

\noindent{\textcolor{blue}{\it Phase diagram.}---} We determine the phase diagram according to DMRG simulations. In the finite $V_1$-$V_2$ region, we find a robust $C=2/3$ FQAH state surrounded by various topologically trivial charge-ordered phases denoted as CDW I-IV with distinct wave vectors specified, as shown in Fig.~\ref{fig_phasediagram}. With small $V_1$ and $V_2$, the ground state is a Fermi Liquid (FL). We show the schematic plots of $n(\mathbf{q})$ for each CDW phase in Fig.\ref{fig_phasediagram}, where $n(\mathbf{q})=\frac{1}{N_x\times N_y}\sum_j e^{-i\mathbf{q}(\mathbf{r_0}-\mathbf{r_j})}(n_j-2/3)$. 

\begin{figure}[htp!]
	\centering		
	\includegraphics[width=0.5\textwidth]{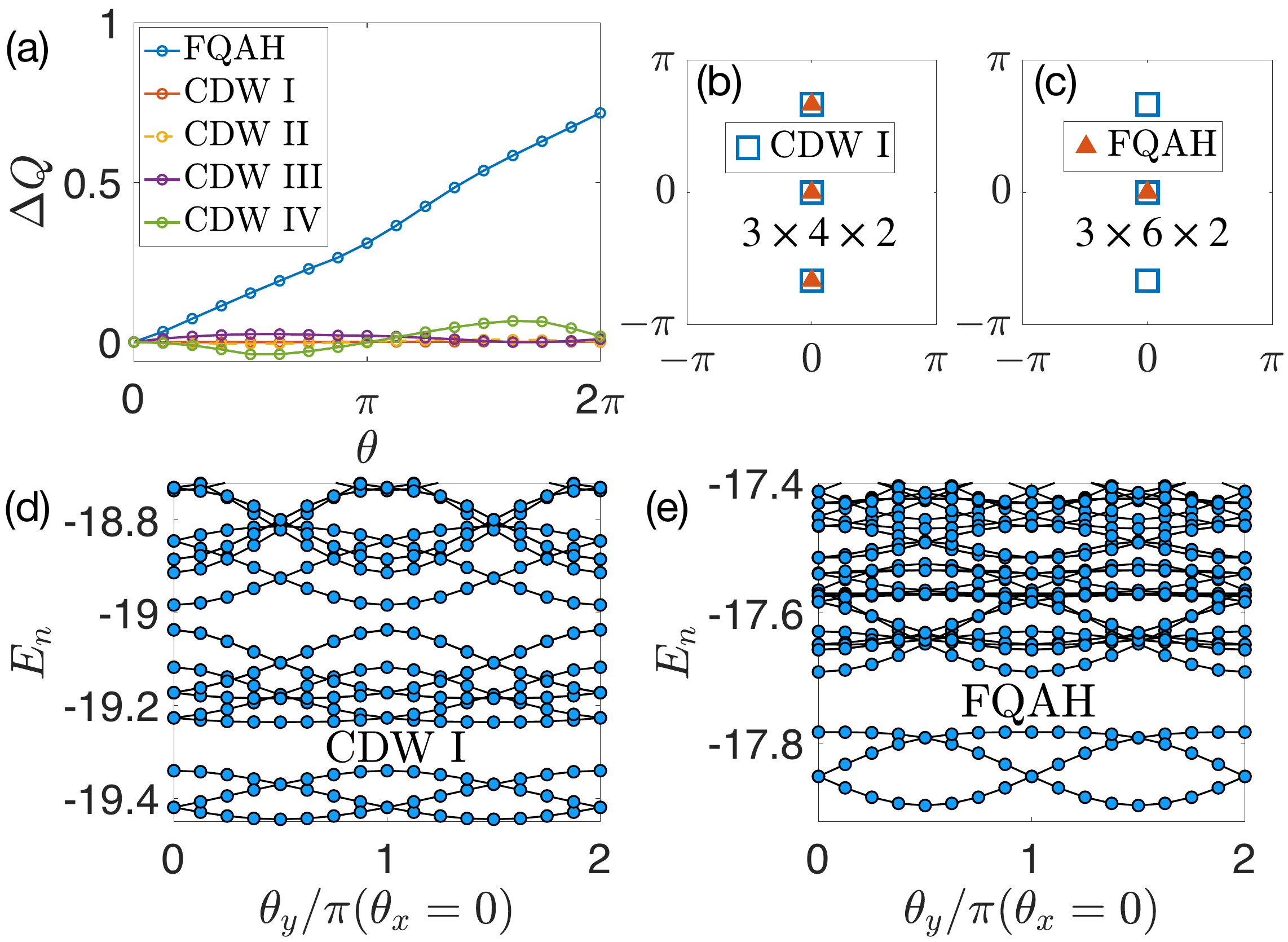}
	\caption{ \textbf{Charge pumping and energy spectra.} (a) We show the DMRG charge pumping results of FQAH ($V_1=1.5, V_2=1$), CDW I ($V_1=2, V_2=0$), CDW II ($V_1=5, V_2=0$), CDW III ($V_1=0, V_2=2$), CDW IV ($V_1=6, V_2=3.2$) states. 
		Unlike the FQAH state, there is no quantized charge pumping in the CDW states. The 3-fold degenerate ground states of CDW I and FQAH states in $4\times4\times2$ and $3\times6\times2$ ED tori are shown in (b) and (c) respectively.  The gapped ED energy spectra of $3\times4\times2$ tori are shown in (d) for CDW I state ($V_1=2, V_2=0$) and (e) for FQAH state ($V_1=2, V_2=1$).}
	\label{fig_TBC}
\end{figure}

We examine the charge pumping of the FQAH state in $N_y=6$ cylinders of DMRG simulations. As we adiabatically introduce a $2 \pi$ flux ($c_i^\dagger c_j^{\ }+h.c. \rightarrow c_i^\dagger c_j^{\ }e^{i\theta}+h.c.$) for hoppings across the periodic boundary in the cylinder, we find two thirds of an electron charge being pumped from one edge of the cylinder to the other, signifying a fractional Hall conductivity $\sigma_{xy}=\tfrac{2}{3} \tfrac{e^2}{h}$ as shown in Fig.~\ref{fig_TBC} (a), while the CDW I-IV states are topologically trivial without quantized charge pumping. 
This is in agreement with our ED results, where the Chern number is $2/3$ for each of the 3-fold degenerate ground state of the FQAH state. Besides, in Fig.~\ref{fig_TBC} (b), in a $3\times4\times2$ torus from ED, the 3-fold degenerate ground states of both FQAH state and CDW I insulator are from different momentum sectors. However, the 3 ground states of FQAH states
in a $3\times6\times2$ torus are all in the $(0,0)$ sector, which satisfies the unique feature of the topological degenracy, in contrast to CDW I states, whose 3-fold ground states of the same geometry are still from different sectors, as shown in Fig.~\ref{fig_TBC} (c). We also show the 3-fold gapped spectra of CDW I and FQAH states with twisted boundary conditions in Fig.~\ref{fig_TBC} (d-e), respectively.

\begin{figure}[htp!]
	\centering		
	\includegraphics[width=0.5\textwidth]{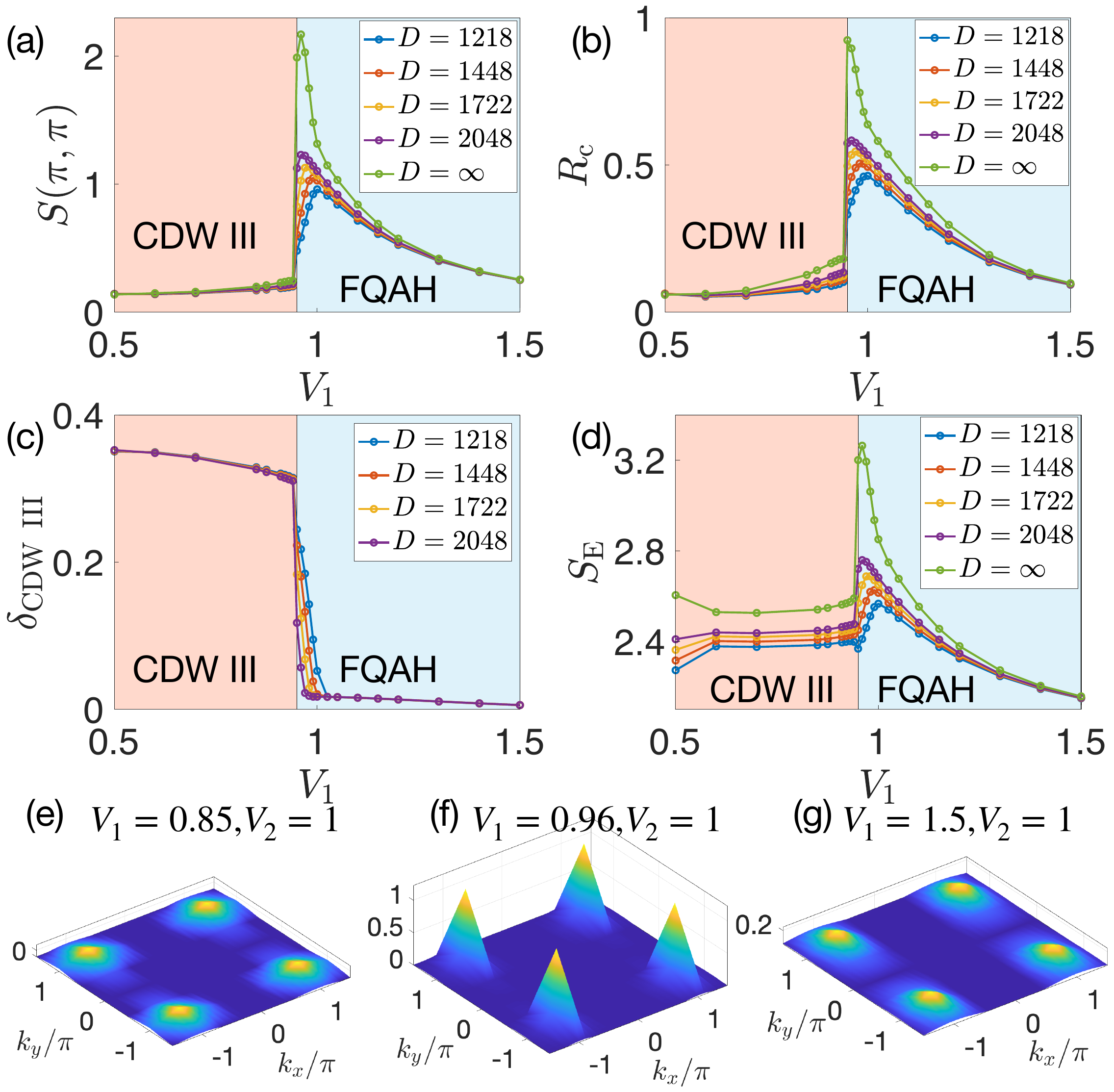}
	\caption{ \textbf{Roton condensation at the FQAH-CDW transition.}  The change of (a) structure factor at $(\pi,\pi)$, (b) correlation ratio, (c) order parameter of CDW III, (d) Bipartite entanglement entropy when cutting the system in the middle, with the change of $V_1$ and fixed $V_2=1$ from DMRG simulations. We also plot the structure factor in the entire Brillouin zone (e) in the CDW III phase, (f) around the transtion point, (g) in the FQAH phase, respectively.}
	\label{fig_roton}
\end{figure}

\noindent{\textcolor{blue}{\it Roton-driven transition.}---} We mainly focus on the FQAH-CDW III transition, where we find the spontaneously translational symmetry breaking in the CDW III state is driven by the condensed roton mode in the FQAH state across their phase transition.
To study this,  we mainly focus on the fixed $V_2=1$ and change $V_1$ on a $4\times24\times2$ cylinder, denoted as the red dotted path in Fig.~\ref{fig_phasediagram}. With small $V_1$, the ground state is CDW III, which is a metallic phase since the CDW order here is incommensurate with filling and cannot induce an insulating phase without topological order, while the FQAH state is in the intermediate $V_1$ region, as shown in Fig.~\ref{fig_phasediagram}.
The change of structure factor, correlation ratio, order parameter, and the entanglement entropy across the FQAH-CDW transition are shown in Fig.~\ref{fig_roton} (a-d), respectively, and we plot the structure factors with maximum bond dimension and different interaction values in Fig.~\ref{fig_roton} (e-g). 
We obtain the structure factor $S(\mathbf{q})=\sum_j e^{-i\mathbf{q}(\mathbf{r_0}-\mathbf{r_j})}( \langle n_0n_j\rangle-\langle n_0 \rangle\langle n_j \rangle)$ by taking the Fourier transformation of density-density correlations. The correlation ratio here is defined as $R_\mathrm{c}=1-\frac{S((\frac{N_x-1}{N_x})\pi,\pi)}{S(\pi,\pi)}$. 

The order parameter of CDW III phase is defined as $\delta_\mathrm{CDW\ III}=|\frac{1}{N_xN_y}\sum (-1)^{i+j}(n^A_{i\mathbf{a_1}+j\mathbf{a_2}}-n^B_{i\mathbf{a_1}+j\mathbf{a_2}})|$, where $\mathbf{a_1}$ and $\mathbf{a_2}$ are the primitive vectors for unit cells and A/B refer to different sublattices (we show the real space patterns of all CDW states in the SM~\cite{suppl}).  The bipartite entanglement entropy is $S_\mathrm{E}=-\sum_i\rho_i\mathrm{ln}\rho_i$ for the reduced density matrix $\rho$.
In CDW III metal phase, $S(\pi,\pi)$ is not so high as shown in Fig.~\ref{fig_roton} (e). 
This is because the DMRG wavefunction turns to suppress the fluctuations in symmetry-breaking phase. Besides, we also find in the FQAH state, the structure factor shown in Fig.~\ref{fig_roton} (g) is similar to that in Fig.~\ref{fig_roton} (e), and this refers to the gapped roton mode at ($\pi,\pi$) with broad dispersion in topological ordered phase. The interesting thing is that when approaching the transition point from FQAH side, $S(\pi,\pi)$ gradually enhances and the correlation ratio is gradually increasing as well (Fig.~\ref{fig_roton} (b)), and these values along with the entanglement entropy (Fig.~\ref{fig_roton} (d)) tend to diverge around the transition point. 
This process is due to the condensation of the roton mode at ($\pi,\pi$), with the finite-momentum charge-neutral excitation gap therein closes. Therefore, around the transition point, the structure factor is of diverging value and quite sharp with $R_\mathrm{c}$ approaching 1. However, the charge distribution is still uniform when the density-density fluctuation goes up quickly (driven by the roton mode), as shown in Fig.~\ref{fig_roton} (c). 
The almost diverging CDW susceptibility suggests this roton-driven transition is either continuous or weakly first-order. Since our DMRG simulations only optimize the lowest state, we thus conclude that the ground-state simulations of structure factors could reveal the hidden information of the excited states in this roton-driven phase transition.

\begin{figure}[htp!]
	\centering		
	\includegraphics[width=0.5\textwidth]{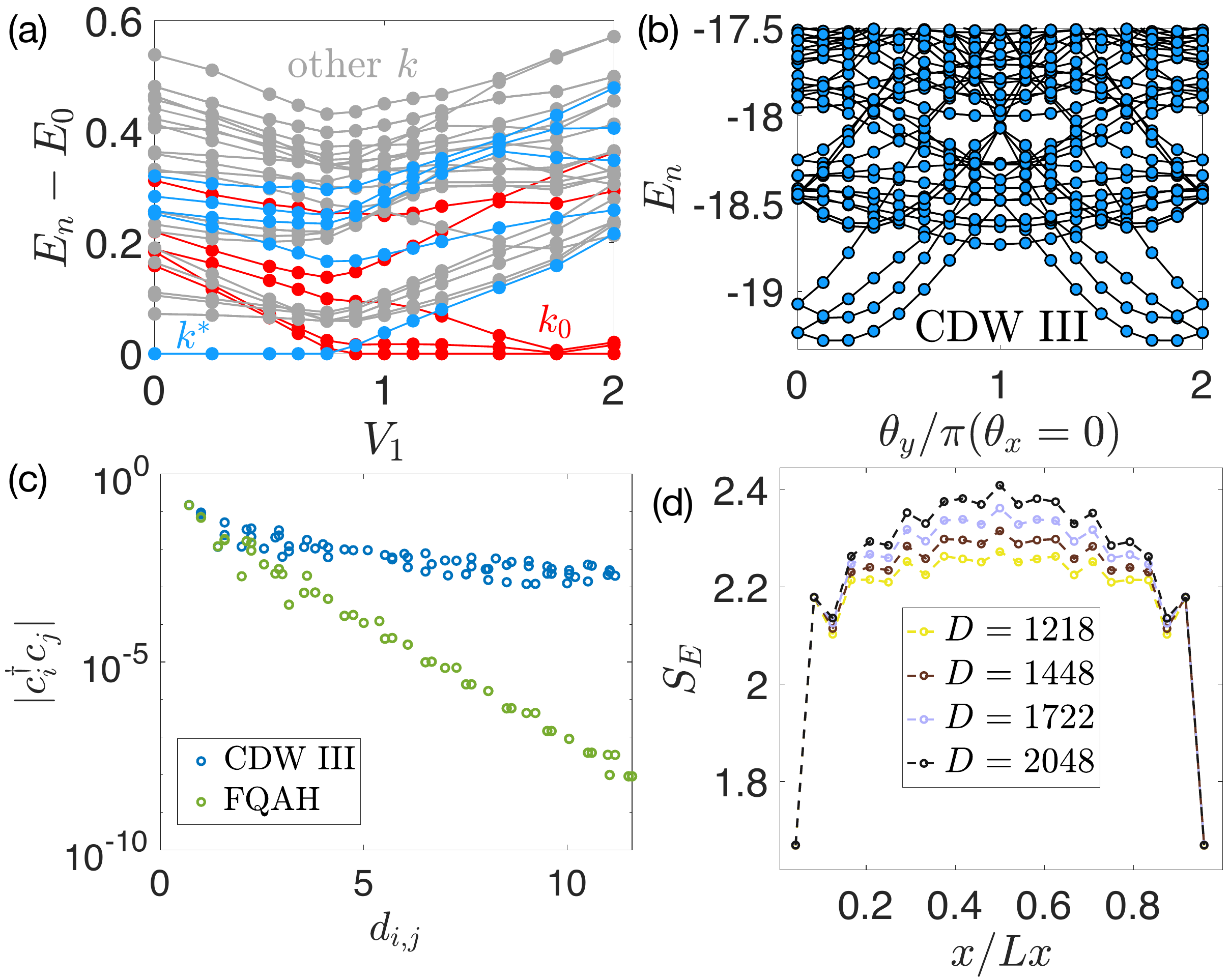}
	\caption{\textbf{Energy spectra and the CDW III metal.}  (a) Energy spectra of a $3\times3\times4$ torus with the change of $V_1$, where we plot the energies of $k_0$ sector in red, $k^\ast$ sectors  in blue, and other momentum sectors in grey. 
		(b) Gapless spectrum of CDW III metal on a $3\times2\times4$ torus with twisted boundary conditions at $V_1=0$ and $V_2=2$. DMRG results showing (c) the decay of correlation function for CDW III ($V_1=0.5$, $V_2=1$) and FQAH ($V_1=1.5$, $V_2=1$) and (d) bipartite entanglement entropy of CDW III metal at $V_1=0.5$ and $V_2=1$ in a $4\times24\times2$ cylinder.}
	\label{fig_ed_transition}
\end{figure}

In Fig.~\ref{fig_ed_transition} (a), we show the energy spectra with varying $V_1$. In the FQAH regime on the $3\times3\times4$ torus in ED, the 3-fold degenerate ground states are all in the $k_0$ sector (which is $(0,0)$). For the CDW III metal, as the CDW order is actually a checkerboard order, the ground-state manifold is 4. If we label the sectors of CDW III ground states as $k^\ast$ (the ground-state sector is determined by the finite-size geometry for this metallic state), we find the lowest state of the $k^\ast$ sector is actually the lowest excited state of the FQAH state. 
This is in agreement with our analysis of DMRG results that the roton minima of the FQAH state in our phase diagram is at ($\pi,\pi$), and the FQAH-CDW III transition is driven by this roton mode, where the momentum transfer of the lowest neutral excitation is constant and this roton gap gradually closes until the $k^\ast$ energy is lower than the $k_0$ energy. 

To further exhibit the metallic nature of CDW III state, we plot the gapless spectrum with twisted boundary conditions in a $3\times2\times4$ torus, power-law decay of correlation function (compared to the exponential decay in the FQAH state) and bipartite entanglement entropy of a $4\times24\times2$ cylinder in Fig.~\ref{fig_ed_transition} (b-d)

\begin{figure}[htp!]
	\centering		
	\includegraphics[width=0.5\textwidth]{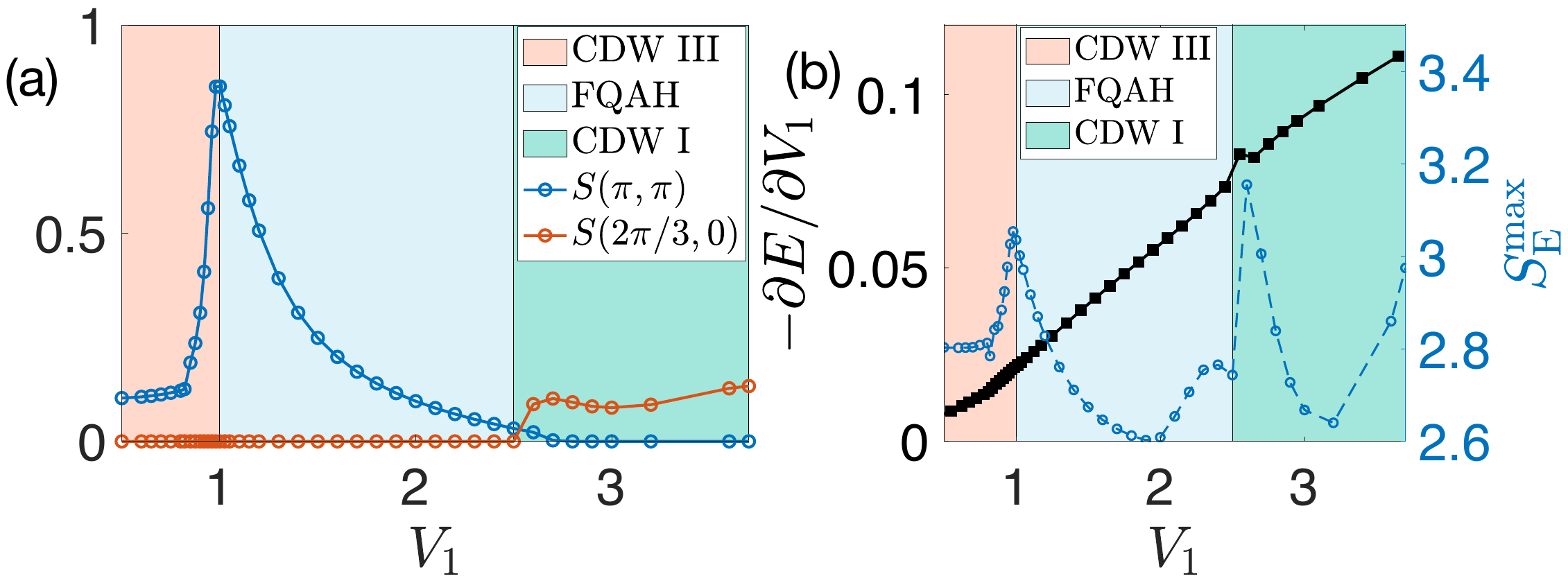}
	\caption{ \textbf{Different FQAH-CDW transitions.}  (a) DMRG results of the change of structure factors at $(\pi,\pi)$ and $(2\pi/3,0)$ with changing $V_1$ and fixed $V_2=1$ on $4\times18\times2$ cylinders. Compared with the CDW III--FQAH transition, the FQAH-- CDW I transition is clearly first order. (b) Corresponding first derivative of energy and maximum entanglement entropy.}
	\label{fig_2transition}
\end{figure}

\noindent{\textcolor{blue}{\it Other FQAH-CDW transitions.}---} 
We also compare the FQAH-CDW III transition and the transition from the FQAH state to the CDW I insulator whose order is different from the roton minima of FQAH states. We plot the change of structure factors at both ($\pi,\pi$) and ($2\pi/3, 0$) in Fig.~\ref{fig_2transition} (a) and the first derivative of energy and maximum entanglement entropy around the bulk of cylinder in Fig.~\ref{fig_2transition} (b). In contrast to the roton-driven transition discussed before, in the FQAH rigime, $S(2\pi/3,0)$ is always $0$, revealing that before the direct transition, there exists no density-density fluctuation at $(2\pi/3,0)$ and there is no roton mode going soft or gradually condense in this transition. The first-order transition nature is also exhibited from the discontinuous first derivative of energy at the FQAH-CDW I transtion point.
 
\noindent{\textcolor{blue}{\it Discussions.}---}  
In FQAH systems, the emergence of a second-order (or weak-first-order) CDW transition from roton condensate shall in general unfolds via one of two scenarios: (1) Should the emerging CDW state be insulating, it is characterized as a FQAH crystal state, meaning it retains fractional topological order alongside CDW order. In such cases, the Hall conductivity is expected to remain the same as in the original FQAH state preceding the roton condensation, because near a second-order phase transition, CDW order is weak and can be treated as a perturbation to the FQAH topological states. (2) In the event that the CDW state is compressible, the resulting state is gapless and the Hall conductivity becomes unquantized. Our model studies provide an illustration of the second scenario. While the first scenario --- a quantum critical point towards a topologically nontrivial insulating CDW state --- will be introduced in our upcoming work~\cite{luFQAH_FQAHS_2024}.

\begin{acknowledgments}
{\it Acknowledgments}\,---\, We thank Bo Yang, Jie Wang and Wei Zhu for helpful discussions. HYL, BBC and ZYM acknowledge the support from the Research Grants Council (RGC) of Hong Kong Special Administrative Region of China (Project Nos. 17301721, AoE/P-701/20, 17309822, HKU C7037-22GF, 17302223), the ANR/RGC Joint Research Scheme sponsored by RGC of Hong Kong and French National Research Agency (Project No. A\_HKU703/22). We thank HPC2021 system under the Information Technology Services and the Blackbody HPC system at the Department of Physics, University of Hong Kong, as well as the Beijng PARATERA Tech CO.,Ltd. (URL: https://cloud.paratera.com) for providing HPC resources that have contributed to the research results reported within this paper. H.Q. Wu acknowledge the support from Guangzhou Basic and Applied Basic Research Foundation (No. 202201011569). The ED calculations reported were performed on resources provided by the Guangdong Provincial Key Laboratory of Magnetoelectric Physics and Devices, No. 2022B1212010008.
\end{acknowledgments}

\bibliographystyle{apsrev4-2}

\newpage\clearpage
\renewcommand{\theequation}{S\arabic{equation}} \renewcommand{\thefigure}{S%
\arabic{figure}} \setcounter{equation}{0} \setcounter{figure}{0}


\begin{widetext}

\section{Supplemental Materials for \\[0.5em]
	Interaction-driven Roton Condensation in a $C=2/3$ Fractional Quantum Anomalous Hall State}
Here, we provide supplementary information of ED simulations in Sec. I, more details about the CDW states in Sec. II, results of other transition path in Sec. III, and details of data extrapolation in Sec. IV.

\subsection{Section I: Supplementary information of ED}
In the main text, we mention that we have used two kinds of real-space geometries. The first kind is similar to the DMRG case, but a torus, where a unit cell of two lattice sites are repeated in space, so we call this $N_y\times N_x\times2$. The other is composed of a square of 4 NN sites, so we call $N_y'\times N_x'\times 4$. We show the momentum sectors of the tori that we have used in Fig.\ref{figsm_BZ}. The reason we use the second kind of geometry is that we need the ($\pi,\pi$) point to study the CDW III state. 

\begin{figure}[htp!]
	\centering		
	\includegraphics[width=0.55\textwidth]{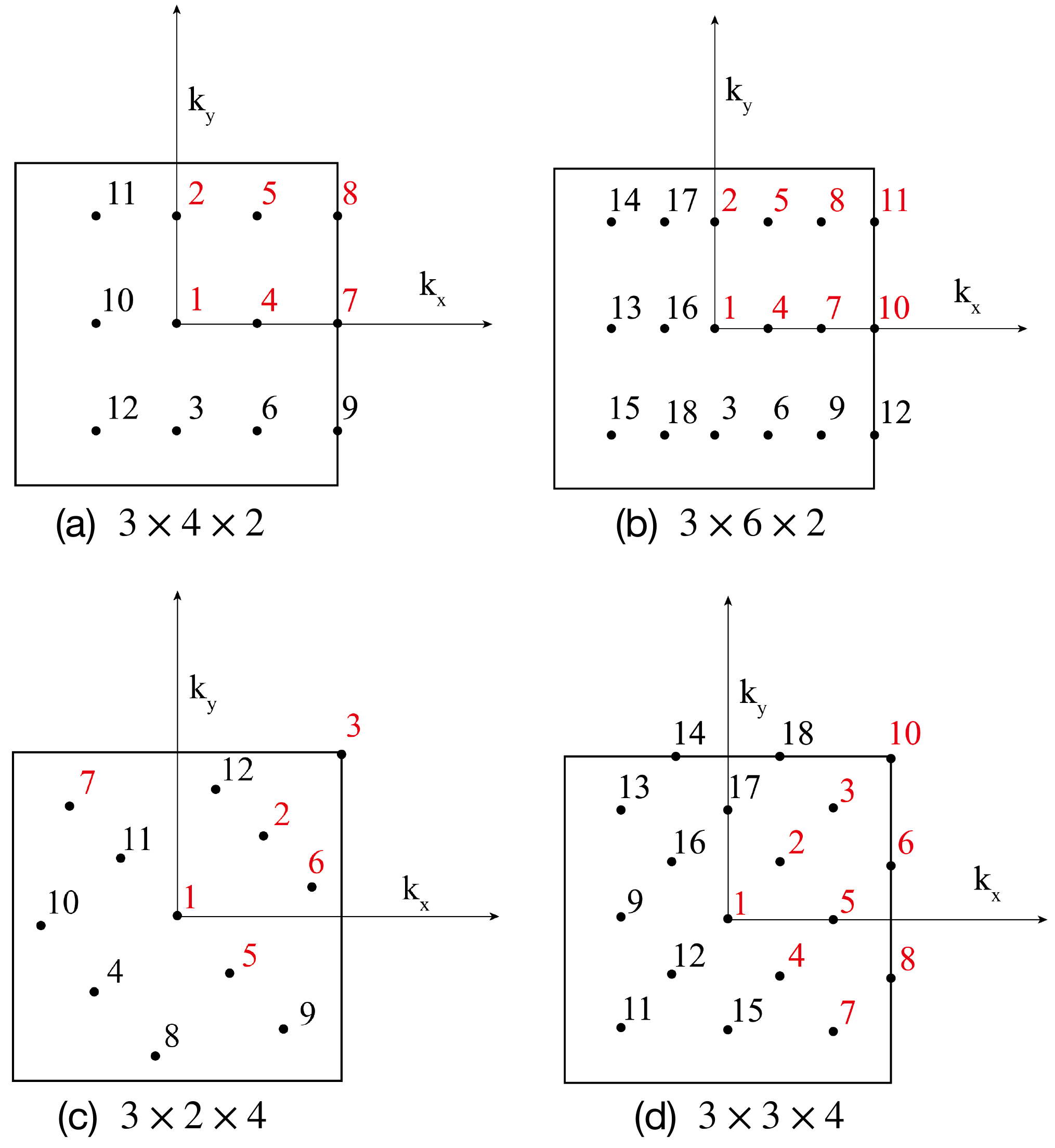}
	\caption{  Corresponding momentum points in the Brillioun zone of unit cells for tori with different real-space geometries. The energy spectra with red momentum sectors are obtained using Lanczos method, while the black ones are obtained via mirror or rotational symmetry.}
	\label{figsm_BZ}
\end{figure}

\begin{figure}[htp!]
	\centering		
	\includegraphics[width=0.3\textwidth]{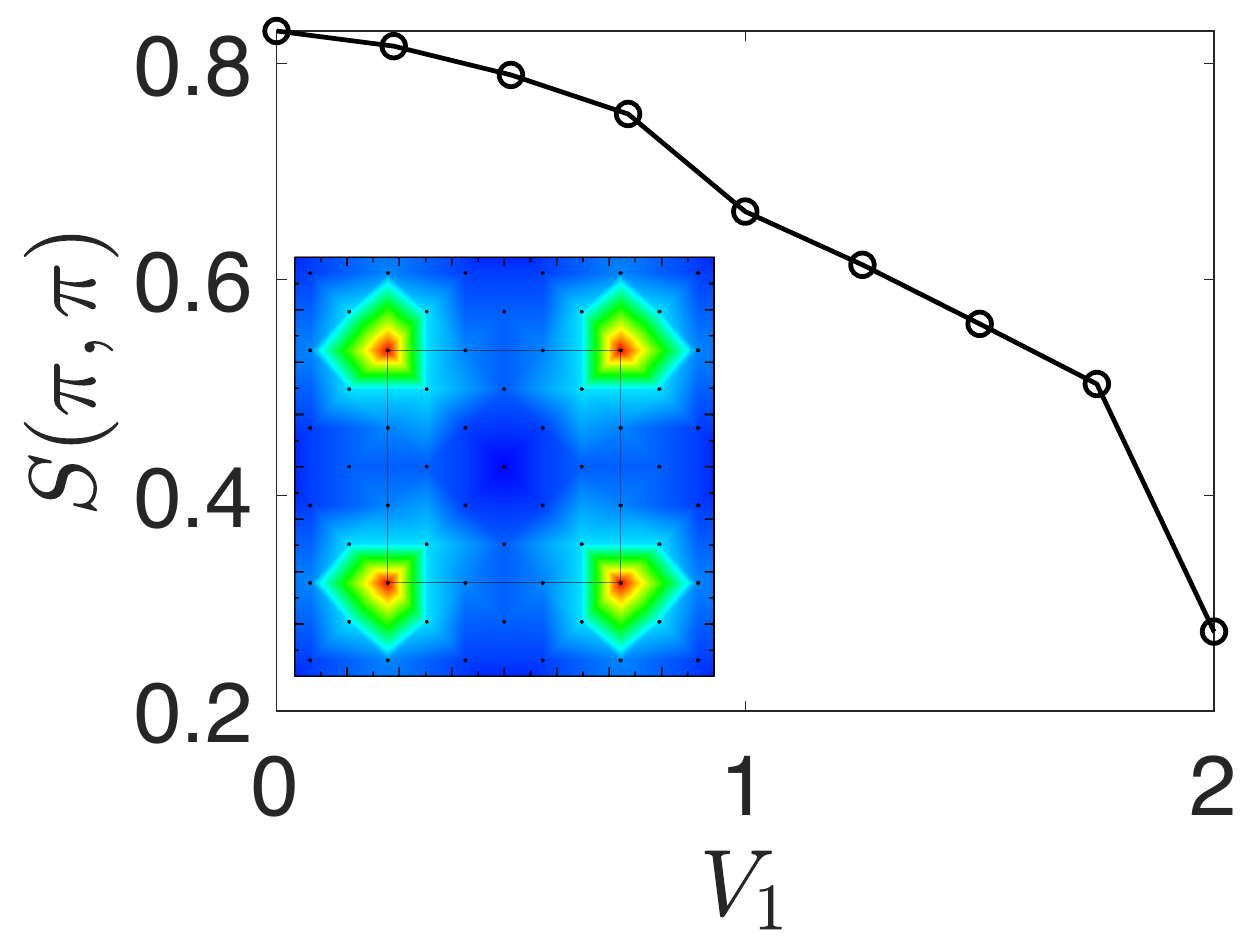}
	\caption{The change of structure factor $S(\pi,\pi)$ with changing $V_1$ and fixed $V_2=1$ on a $3\times\times4$ torus, while the energy spectrum on the same path is shown in Fig.4(a). The inset is the plot of structure factor with the peaks at ($\pi,\pi$). }
	\label{figsm_ed_spipi}
\end{figure}
\newpage
The ED spectra of the FQAH-CDW III transition is shown in Fig.4(a). Since the translational symmetry is not explicitly broken in the ED simulations, the structure factors here actually refer to the strength of CDW order but not susceptibility, and the diverging behavior of $S(\pi,\pi)$ at transition point in the ground-state result will not be exhibited as in the DMRG simulations, as shown in Fig.\ref{figsm_ed_spipi}.

\subsection{Section II: Details of CDW I-IV states}
In Fig.1, the global $V_1-V_2$ phase diagram is given and the schematic plots of $n(\mathbf{q})$ (Fourier transformation of density occupation numbers) of the CDW states are shown. Besides, the trivial topology of CDW I-IV is shown in Fig.2 (a). Here, we show more complementary information of the CDW I-IV states.

\begin{figure}[htp!]
	\centering		
	\includegraphics[width=0.6\textwidth]{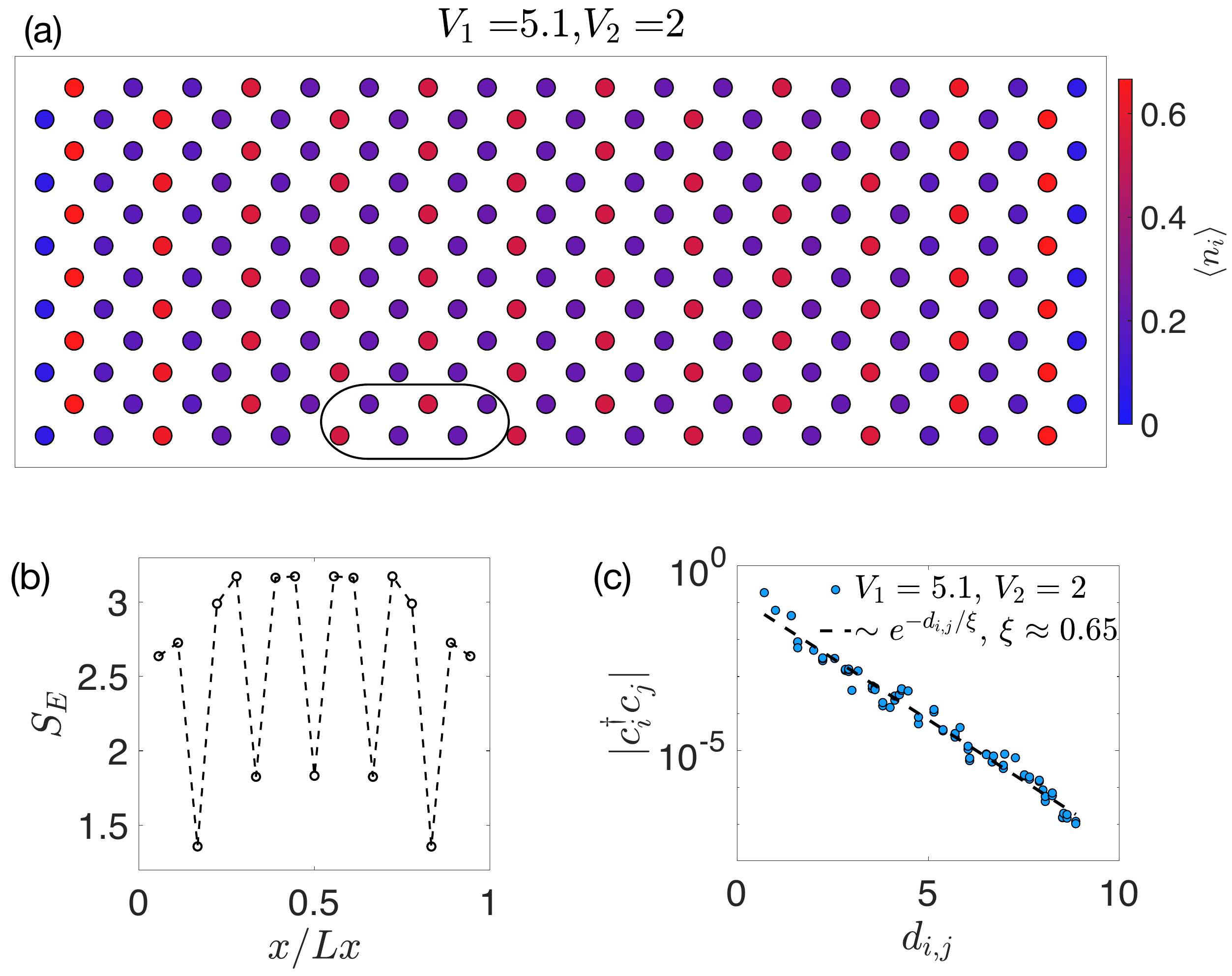}
	\caption{  (a) Real-space charge distribution of CDW I state on a $N_y=6$ cylinder with $V_1=5.1$ and $V_2=2$. The black ellipse represents the enlarged unit cell and the minimal effective cell. (b) Corresponding bipartite entanglement entropy and (c) Exponential decay of correlation function with the reference site in the center of cylinder.}
	\label{figsm_CDWI}
\end{figure}

First, we show the real-space charge patterns of CDW I-state in Fig.\ref{figsm_CDWI} (a), which corresponds to the $(2\pi/3,0)$ order. The entanglement entropy with a plateau in the bulk of cylinder and the exponential decay of the correlation function further exhibit the insulating behavior of CDW I state, as shown in Fig.\ref{figsm_CDWI}(b,c).

\newpage
\begin{figure}[htp!]
	\centering		
	\includegraphics[width=0.7\textwidth]{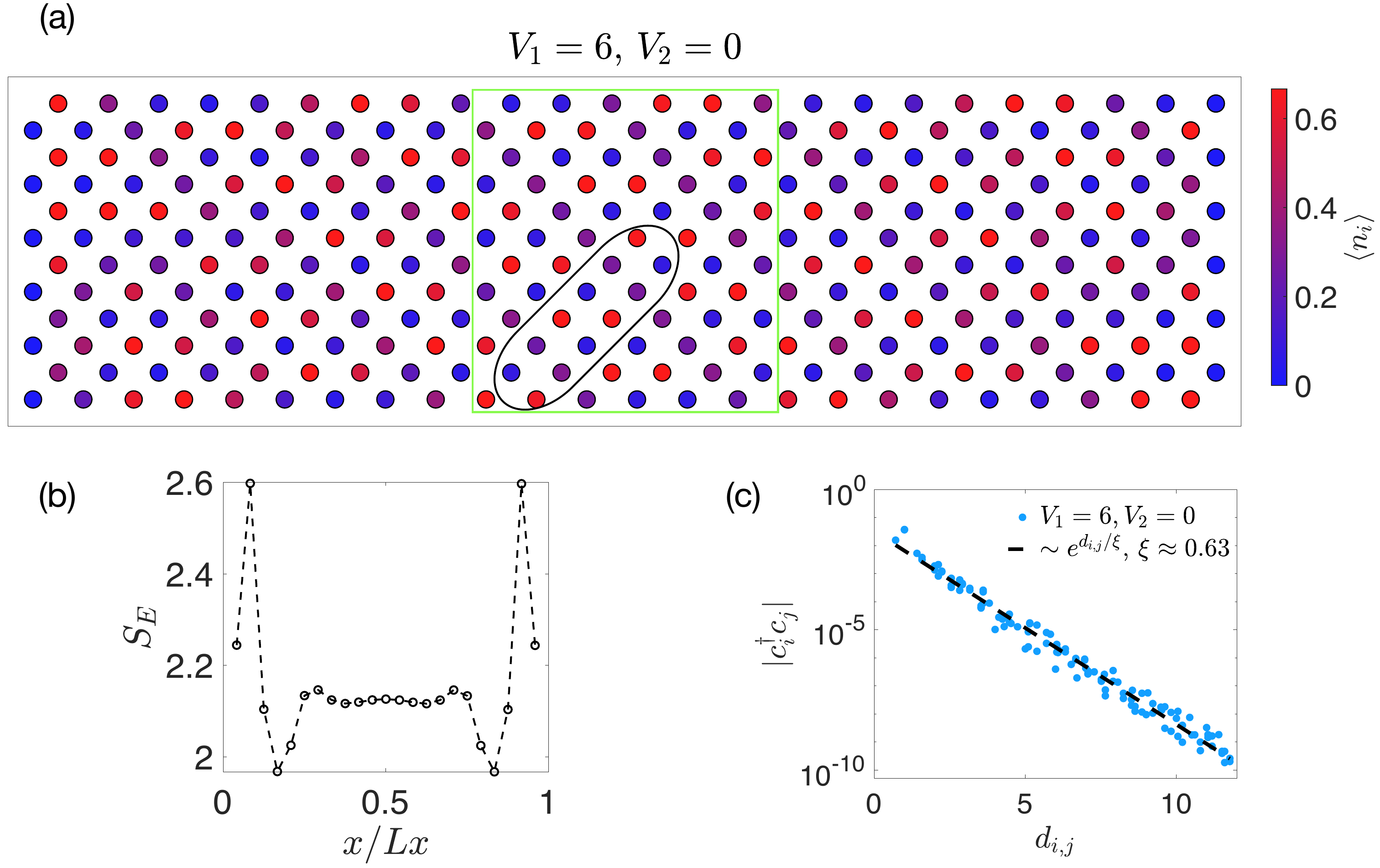}
	\caption{  (a) Real-space charge distribution of CDW II state on a $N_y=6$ cylinder with $V_1=6$ and $V_2=0$. The black ellipse represents the minimal effective cell while the green square refer to the enlarged periods along the original primitive vectors. (b) Corresponding bipartite entanglement entropy and (c) Exponential decay of correlation function with the reference site in the center of cylinder.}
	\label{figsm_CDWII}
\end{figure}

For CDW II, we show the real-space charge patterns in Fig.\ref{figsm_CDWII} (a), which corresponds to the $(\pi/3,\pi/3)$ order. According to the bipartite entanglement entropy and correlation functions with exponential decay behavior up to the present maximum bond dimension, this commensurate CDW II state is very likely an insulator.

\begin{figure}[htp!]
	\centering		
	\includegraphics[width=0.6\textwidth]{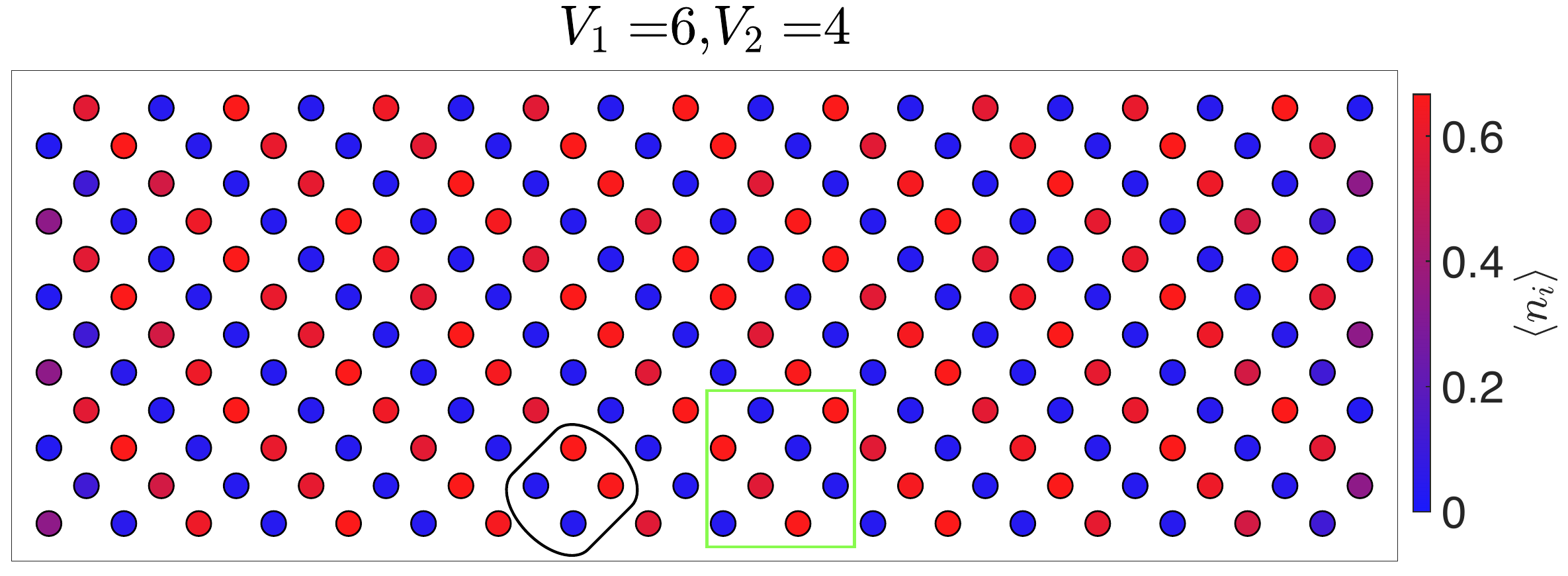}
	\caption{   Real-space charge distribution of CDW III state on a $N_y=6$ cylinder with $V_1=6$ and $V_2=4$. The black ellipse represents the minimal effective cell while the green square refer to the enlarged periods along the original primitive vectors.}
	\label{figsm_CDWIII}
\end{figure}

\begin{figure}[htp!]
	\centering		
	\includegraphics[width=0.6\textwidth]{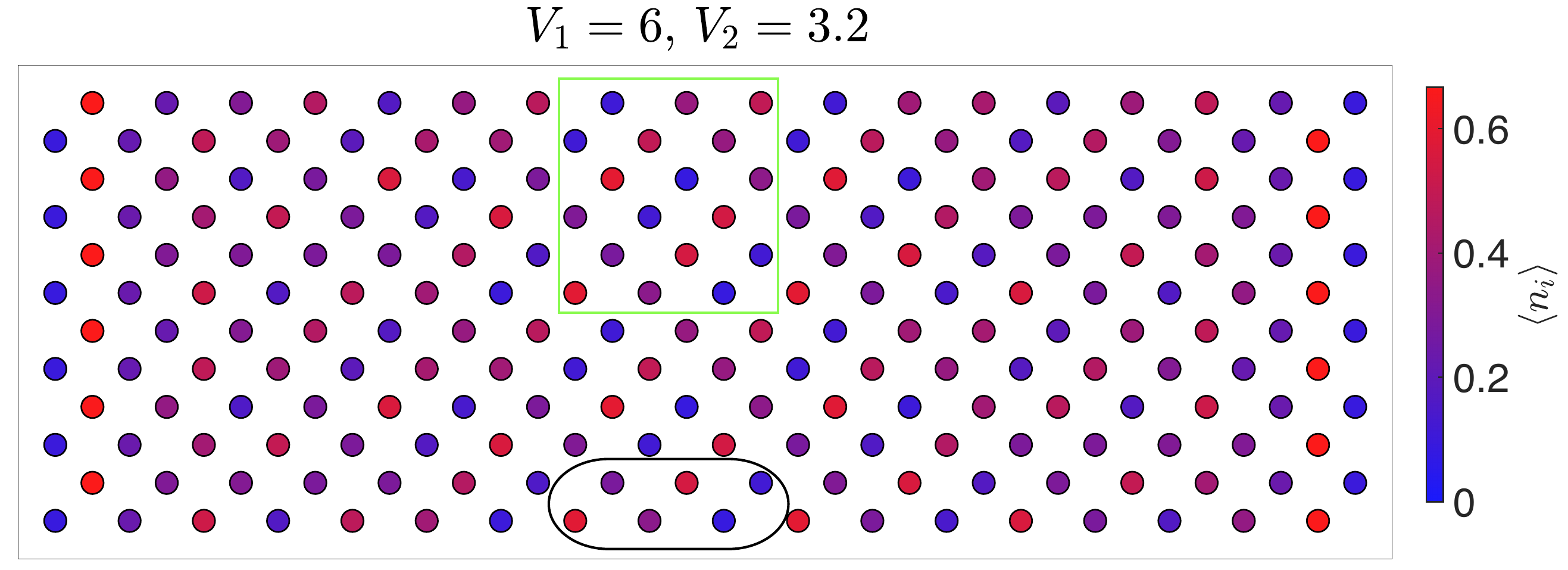}
	\caption{   Real-space charge distribution of CDW IV state on a $N_y=6$ cylinder with $V_1=6$ and $V_2=3.2$. The black ellipse represents the minimal effective cell while the green square refer to the enlarged periods along the original primitive vectors.}
	\label{figsm_CDWIV}
\end{figure}

As the details of CDW III metal has been well discussed in the main text, here we show the real-space charge pattern at $V_1=6$ and $V_2=4$ in Fig.\ref{figsm_CDWIII}. We also plot the charge distribution of CDW IV in Fig.\ref{figsm_CDWIV}, since the parameter space is small in the considered phase diagram and the transitions related to this state is not the focus in this work, we leave the more detailed analysis of this state for future study.

\subsection{Section III: Other phase transitions }

In the main text, we have exhibited the roton-triggered FQAH-CDW III transition with $V_2=1$, and here we show such a transition along another path with $V_2=2$ in Fig.\ref{figsm_transition_v2_2}.

\begin{figure}[htp!]
	\centering		
	\includegraphics[width=0.6\textwidth]{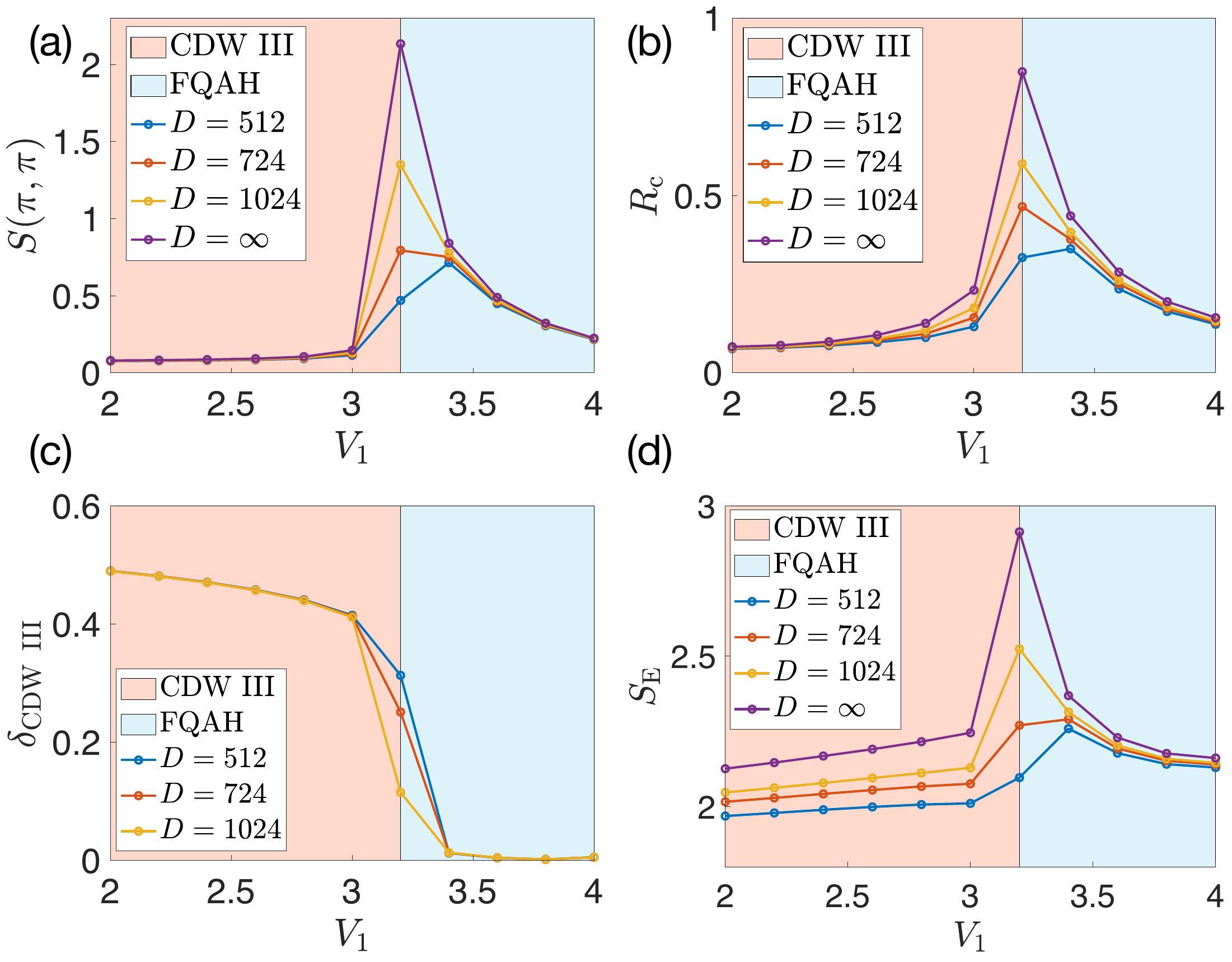}
	\caption{The change of structure factor, correlation ratio, order parameter of CDW III, and bipartite entanglement entropy with the change of $V_1$ and fixed $V_2=2$ on a $4\times18\times2$ cylinder.}
	\label{figsm_transition_v2_2}
\end{figure}

It is very similar to the case introduced in the main text, as we see this FQAH-CDW III transition is driven by the $(\pi,\pi)$ roton mode. The strcture factor $S(\pi,\pi)$ increases fast and becomes sharper in FQAH state when the order has not been established, and they tend to diverge at the transition point, together with the entanglement entropy, which refer to the closing roton gap with $\mathbf{q}=(\pi,\pi)$. We note that, even with lower bond dimension, the transition here with fixed $V_2=2$ seems more continuous than that of fixed $V_1=1$, with higher $S(\pi,\pi)$ and correlation ratio. Therefore, it would be interesting in the future work to simulate more transition points between FQAH and CDW III to find the most continuous one at the finite size, which might be beneficial to possible analysis of the quantum critical point of the roton-driven transition.

\newpage
\subsection{Section IV: Details of bond-dimension extrapolations}
\begin{figure}[htp!]
	\centering		
	\includegraphics[width=0.8\textwidth]{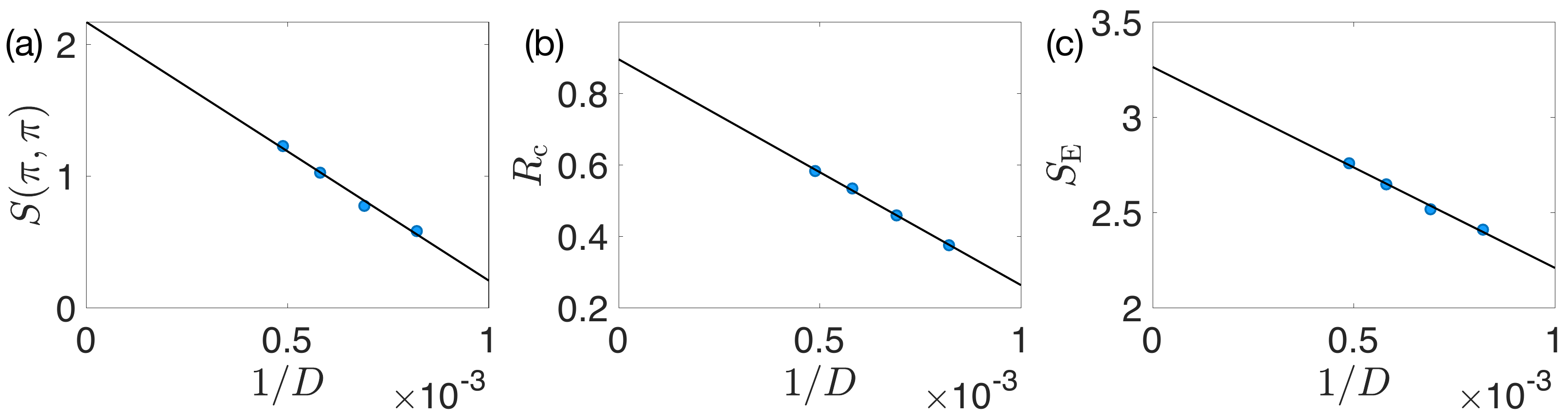}
	\caption{The extrapolation of (a) $S(\pi,\pi)$, correlation ratio, and bipartite entanglement entropy to $1/D$ at $V_1=0.96$ and $V_2=1$, where $D$ is the bond dimension.}
	\label{figsm_extrap}
\end{figure}
In Fig.3, we have extrapolated the data with $D=1218, 1448, 1722, 2048$. Here, we show the details of extrapolation and take the data at $V_1=0.96$ and $V_2=1$ as an example in Fig.\ref{figsm_extrap}. The diverging structure factor, correlation ratio (approacing 1), and entanglement entropy are well fitted to the scaling of $1/D$.

\end{widetext}
\end{document}